\documentclass[letterpaper,twocolumn,american,showpacs,pra,aps]{revtex4}
\usepackage{amsmath}
\usepackage{graphicx}
\usepackage{amssymb}
\usepackage{subfigure}
\newbox\subfigbox
\makeatletter
\newenvironment{subfloat}%
{\def\caption##1{\gdef\subcapsave{\relax##1}}%
  \let\subcapsave\@empty%
  \setbox\subfigbox\hbox%
  \bgroup}%
{\egroup%
  \subfigure[\subcapsave]{\box\subfigbox}}%
\makeatother
\def\<{\begin{equation}}
\def\>{\end{equation}}

\begin{document}
\title{ Characterization of topological states on a lattice with Chern number}
\author{Mohammad Hafezi$^{1}$, Anders S. S\o rensen$^{2}$, Mikhail D. Lukin$^{1}$, Eugene Demler$^{1}$}
\affiliation{$^1$ Physics department,  Harvard university, Cambridge
Massachusetts 02138 and \\
$^2$ QUANTOP, Danish National Research Foundation Centre of Quantum
Optics, Niels Bohr Institute, DK-2100 Copenhagen \O, Denmark }

\begin{abstract}

We study Chern numbers to characterize the ground state of strongly
interacting systems on a lattice. This method allows us to perform a
numerical characterization of bosonic fractional quantum Hall (FQH)
states on a lattice where conventional overlap calculation with
known continuum case such as Laughlin state, breaks down due to the
lattice structure or dipole-dipole interaction. The non-vanishing
Chern number indicates the existence of a topological order in the degenerate ground state manifold.
\end{abstract}


\maketitle

One of the most dramatic manifestations of interactions in many-body
systems is the appearance of new quantum states of matter. Often such states
can be characterized by an order parameter and spontaneous breaking
of some global symmetries. The `smoking gun' evidence of such states
is the appearance of Goldstone modes of spontaneously broken
symmetries. However, some types of quantum many-body phases can not
be characterized by a local order parameter. Examples can be found
in FQH systems\cite{girvin87}, lattice gauge theories\cite{lee06},
and spin liquid states\cite{anderson}.  Such states can be
characterized by the topological order\cite{wen95} which encompasses
global geometrical properties such as ground state degeneracies on
non-trivial manifolds\cite{wen90}. Topologically ordered states
often exhibit fractional excitations\cite{senthil} and have been
proposed as a basis of a new approach to quantum
computations\cite{kitaev03}. However, in many cases, identifying a
topologically ordered state is a challenging task even for
theoretical analysis. Given an exact wavefunction of the ground
state in a finite system, how can one tell whether it describes a
FQH phase of a 2D electron gas or a spin liquid phase on a lattice?
One promising direction to identifying topological order is based on
the Chern number calculations\cite{hatsugai05}. The idea of this
approach is to relate the topological order to the geometrical phase
of the many-body wavefunction under the change of the boundary
conditions\cite{niu85}.

Important work of  Berry \cite{berry} and Simon \cite{simon}
initiated the investigation on geometrical phase factors and since
then the field has been extensively studied in different contexts --
for a review see, for example, \cite{geo_phase}. In quantum Hall
(QH) systems, early works on the Chern number analysis \cite{kohmoto} is focused on
the Hall conductance and robustness of QH states against changes in the band
structure\cite{TKNdN} and the presence of
disorder\cite{wen90,Sheng03}.  Currently, there is also considerable
interest in understanding FQH states in the presence of a strong
periodic potential.  Such systems are important in several contexts
including anyonic spin states \cite{kitaev2005}, vortex liquid
states \cite{balents05}, and ultracold atoms in optical
lattices\cite{sorensen, hafezi, Jaksch06,mueller, jaksch} which are promising
candidates for an experimental realization.

In this letter, we investigate a novel procedure for calculating
Chern numbers and demonstrate that this method provides insight into
the topological order of the ground state in regimes where other
methods fails to provide a definite answer for the nature of the
ground state wavefunction. In particular, we study a fractional
quantum Hall system with bosons on a lattice with a filling factor
of $\nu=1/2$, where $\nu$ is the ratio of the number of magnetic
flux quanta to the number of particles. In the continuum limit,
where the flux-fraction through each plaquette $\alpha$ is very
small ($\alpha \ll 1$), this system is exactly described by the
Laughlin wavefunction. However, in a recent study \cite{sorensen},
it was found that for some values of $\alpha \gtrsim 0.25$, the
Laughlin wavefunction ceases to be a good description of the system,
indicated by a decreasing overlap between the ground state and the
Laughlin wavefunction. From this study, it is unclear whether this
represented a change in the nature of the ground state, or just that
the lattice structure distorts the state. Here, we use the Chern
number calculation to provide an unambiguous characterization of the
ground state even outside the regime where there is a significant
overlap with the Laughlin wavefunction. In particular, we show that the Chern
number and hence the topological order of the system remains
undisturbed until $\alpha \lesssim 0.4$.(Tab.\ref{table:chern}).

\begin{figure}
  \centering
    \begin{subfloat}
    \label{twist}
    \includegraphics[width=.30 \textwidth]{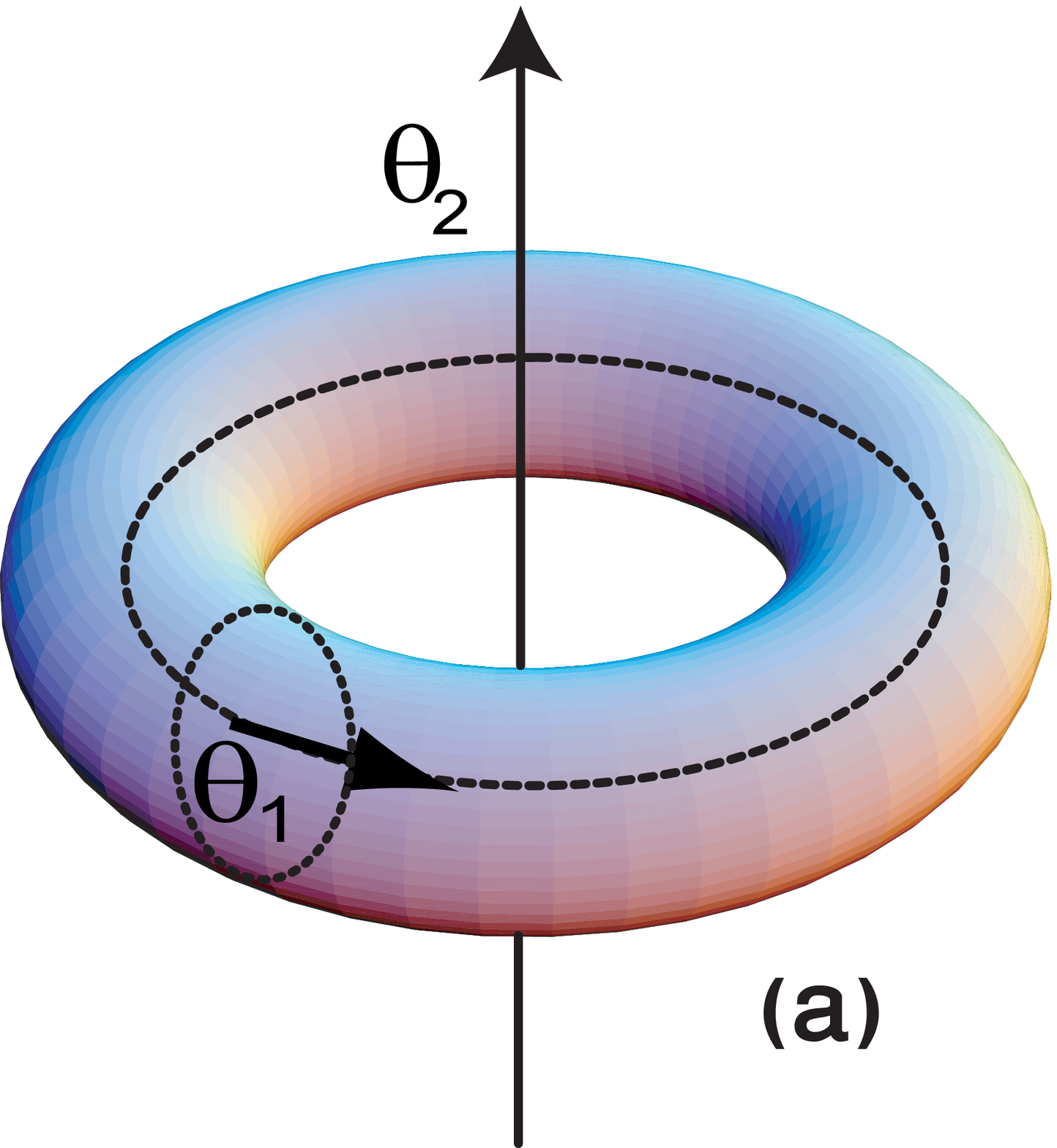}
    \end{subfloat}
 \hspace{0cm}
 \centering
    \begin{subfloat}
    \label{gauge}
    \includegraphics[width=.30 \textwidth]{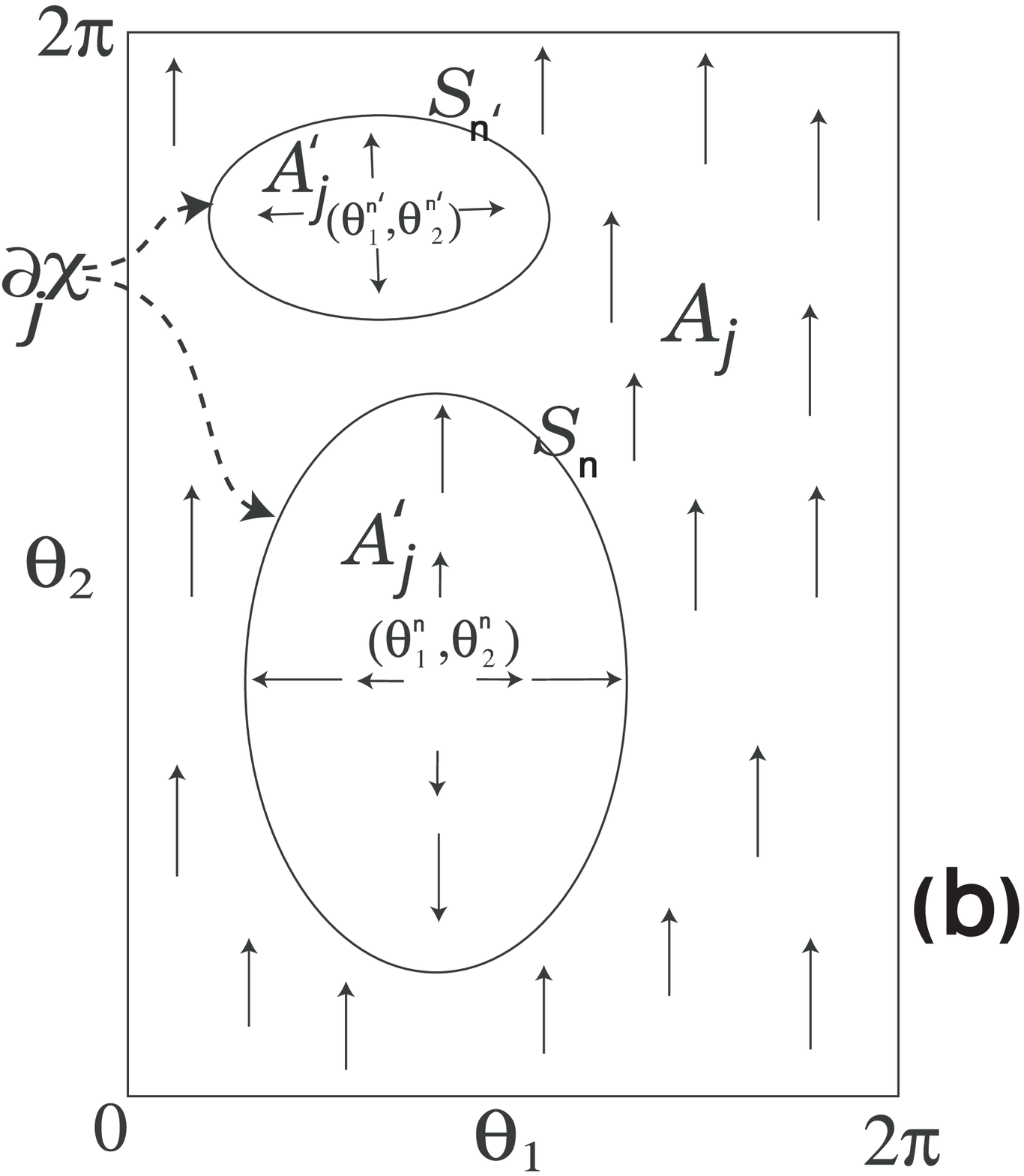}
    \end{subfloat}

\caption{ (a) Twist angles of the boundary condition as the result
of two magnetic fluxes threaded the torus (b) Redefining the vector
potential around the singularities: $\mathcal{A}_j$ is not
well-defined everywhere on the torus of the boundary condition.
Therefore, another vector field $\mathcal{A'}_j$ with different
definition should be introduced around each singularity
$(\theta_1^n,\theta_2^n)$ of $\mathcal{A}_j$. However,
$\mathcal{A}_j$ and $\mathcal{A'}_j$ are related to each other with
a gauge transformation $\chi$ and the Chern number depends only on
the loop integrals of $\chi$ around those singularity regions.}
\label{torus}

\end{figure}

To introduce the basic ideas of the Chern number calculation in
many-body system, we first discuss the continuum regime. We consider
a single particle with charge $\textbf{q}$ on a torus $ T^2(L_x
\times L_y)$ in the presence of a magnetic field $\textbf{B}$
perpendicular to the surface. The corresponding Hamiltonian is
invariant under the magnetic translation of single particle
$\textbf{s}$, $t_s(\textbf{a})=e^{i \textbf{a}\cdot k^s/\hbar}$
where $\textbf{a}$ is a vector on the torus, and $k^s$ is the
pseudo-momentum of particle $\textbf{s}$, defined by $k_{x(y)}= -i
\hbar \frac{\partial}{\partial x(y)}-q A_{x(y)} \mp q By(x)$ in x
and (y) direction, respectively, and $\vec{A}$ is the corresponding
vector potential. The generalized boundary condition is given by the
translation, $ t_s(L_i\hat{i})\psi(x_s,
y_s)=e^{i\theta_i}\psi(x_s,y_s) \label{twist_angles} $, where
$(i=1,2)$ refer to two directions (x,y) on the torus $T^2$ and the
$\theta_i$'s are twist angles of the boundary (Fig.\ref{torus}a).
The magnetic phase through each plaquette ($2 \pi \alpha$) arises
from the field perpendicular to the surface of the torus. The Chern
number for non-degenerate state $\alpha$ is defined by,

\< C(\alpha)= \frac{1}{2 \pi}\int_0^{2\pi} d\theta_1 \int_0^{2\pi}
d\theta_2 (\partial_1 \mathcal{A}_2^{(\alpha)}-\partial_2
\mathcal{A}_1^{(\alpha)}) \label{eq:chern}\> where
$\mathcal{A}_j^{(\alpha)}(\theta_1,\theta_2)$ is defined as a vector
field based on the eigenstate $\Psi^{(\alpha)} (\theta_1,\theta_2)$
on $T^2$ by $ \mathcal{A}_j^{(\alpha)}(\theta_1,\theta_2) \doteq i
\langle\Psi^{(\alpha)}|\frac{\partial}{\partial
\theta_j}|\Psi^{(\alpha)}\rangle \label{eq:vectorfield}$.

In the context of QH systems, the time derivative of twist angles
could be considered as voltage drops across the Hall device in two
dimensions and the boundary averaged Hall conductance of the any
state is related to the Chern number of that state \cite{niu85}:
$\sigma_{H}^\alpha=C(\alpha) e^2/h$.

The non-trivial behavior (non-zero conductance in the case of
quantum Hall system) occurs because of singularities of the vector
field. If for a given non-degenerate state the corresponding vector
field is not defined for certain angles $(\theta_1^n,\theta_2^n)$ in
$S_n$ regions (Fig.\ref{torus}b), then we should introduce a new
well-defined vector field
$\mathcal{A'}_j^{(\alpha)}(\theta_1,\theta_2)$, inside those
regions. These two vector fields differ from each other by a gauge
transformation, $\mathcal{A}_j^{(\alpha)}(\theta_1,\theta_2)-
\mathcal{A'}_j^{(\alpha)}(\theta_1,\theta_2) = \partial_j
\chi(\theta_1,\theta_2) \label{eq:1D_gauge}$ and the Chern number
reduces to the winding number of the gauge transformation
$\chi(\theta_1,\theta_2)$ over small loops encircling
$(\theta_1^n,\theta_2^n)$, i.e. $\partial S_n$ \cite{kohmoto},

\<C(\alpha)= \sum_n \frac{1}{2 \pi} \oint_{\partial S_n}
\overrightarrow{\nabla} \chi \cdot d \overrightarrow{\theta}.
\label{chern_gauge}\>

For the case of degenerate ground state a generalization of the
above argument can be made, where instead of having a single vector
field $ \mathcal{A}_j^{(\alpha)}(\theta_1,\theta_2) $, a tensor
field $ \mathcal{A}_j^{(\alpha, \beta)}(\theta_1,\theta_2) $ should
be defined, $\alpha, \beta= 1 ... q$ for a $q$-fold degenerate
ground state: $\mathcal{A}_j^{(\alpha,\beta)}(\theta_1,\theta_2)
\doteq i \langle\Psi^{(\alpha)}|\frac{\partial}{\partial
\theta_j}|\Psi^{(\beta)}\rangle$.

Therefore, when $\mathcal{A}_j^{(\alpha,\beta)}$ is not defined,
similar to the non-degenerate case, a new gauge convention should be
acquired for those regions with singularities. This gives rise to a
tensor gauge transformation on the border of these regions, $
\partial_j \chi^{(\alpha,\beta)}(\theta_1,\theta_2)=
\mathcal{A}_j^{(\alpha,\beta)}(\theta_1,\theta_2)-
\mathcal{A'}_j^{(\alpha,\beta)}(\theta_1,\theta_2)$ and consequently
the Chern number is given by the trace of the tensor $\chi$,

\< C(1,2,...,q)=\sum_n \frac{1}{2 \pi} \oint_{\partial S_n}
\overrightarrow{\nabla} \textrm{Tr}~\chi^ {(\alpha,\beta)}
 \cdot d \overrightarrow{\theta}. \label{eq:degenerate}\>

We focus on a system of bosons on a square lattice described by the
Hamiltonian \cite{hafezi}:

\begin{eqnarray} H &=&-J \sum_{x,y} \hat{a}^\dag_{x+1,y} \hat{a}_{x,y}e^{-i
\pi \alpha y}+ \hat{a}^\dag_{x,y+1} \hat{a}_{x,y}e^{i \pi \alpha
x}+h.c. \nonumber \\&+& U \sum_{x,y}  \hat{n}_{x,y}(
\hat{n}_{x,y}-1), \label{eq:Hamiltonian}
\end{eqnarray}
 \\
where $J$ is the hopping energy between two neighboring sites, $U$
is the on-site interaction energy, and $2 \pi \alpha$ is the phase
acquired by a particle going around a plaquette. We concentrate on
the hardcore limit ($U \gg J$) and $\nu=1/2$ where $\nu$ is the
ratio of the number of particles to the total number of flux in the
system. The experimental proposal for realizing such a Hamiltonian
for atomic gases confined in an optical lattice has already been
investigated \cite{sorensen, hafezi}. The ground state of the system
for very dilute lattice $\alpha \lesssim 0.2$ is two-fold degenerate
and is well described by Laughlin state. When $\alpha$
increases the lattice structure becomes more apparent and the
overlap with Laughlin wavefunction breaks down. However, by
numerical calculation, we show that Chern number characterizes
system better and remains the same, i.e. 1/2 for each state in the
ground state manifold, for systems with higher flux density $\alpha
\lesssim 0.4$.

\begin{figure}
  \centering

 \includegraphics[width=.50 \textwidth]{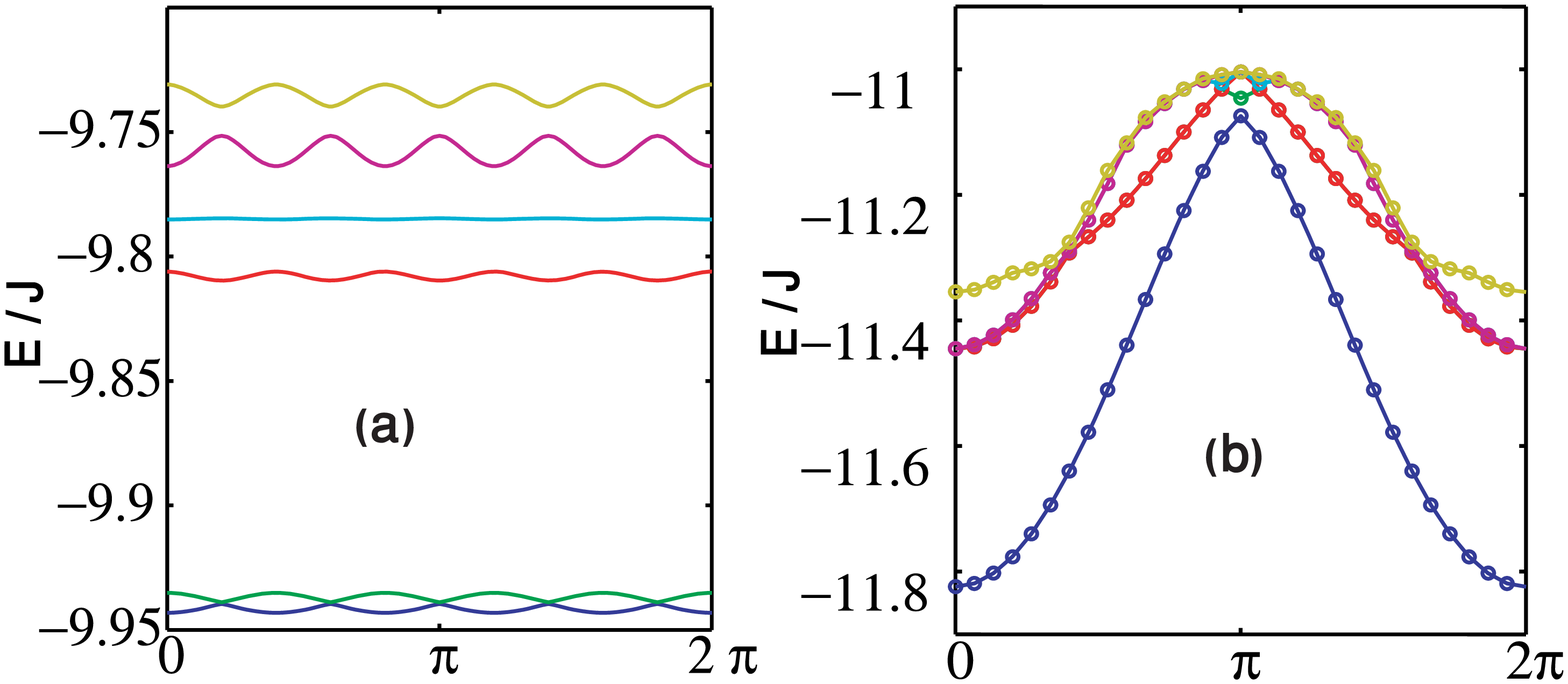}

\caption {Low-lying energy levels as a function of twist angles. For
finite $\alpha$ the ground state energy oscillates as a function of
twist angles and for high $\alpha \gtrsim 0.4$ the oscillations
reach the exited levels. (a) shows lowest energy levels for
$\alpha=0.32 $ (4 atoms on a 5x5 lattice) while (b) shows the five
lowest energy levels for $\alpha= 0.4$ ( 5 atoms on a 5x5 lattice)
In both plots, $\theta_2=\pi$ and $\theta_1$ is varied from zero to
$2 \pi$.}
 \label{fig:energy-levels}

\end{figure}

In the case of a very dilute lattice $\alpha \ll 1$, i.e. the
magnetic length is much larger than the lattice spacing, and hence
the lattice Hamiltonian approaches the continuum limit. According to
Haldane \cite{haldane85}, the magnetic translational symmetry of the
center of mass results in a two-fold degeneracy of the ground state
for $\nu=1/2$. However, by increasing the magnetic field, the
lattice structure becomes more pronounced, in such a way that even
for a single particle, the lattice modifies the energy levels from
being simple Landau levels into  the fractal structure known as the
Hofstadter butterfly \cite{hofstadter}. For the many-body problem,
the presence of the lattice causes the energy levels to oscillate
(see Fig.\ \ref{fig:energy-levels}a) and instead of having a unique
degeneracy for all twist angles values, the ground state is two-fold
degenerate at only certain twist angles. However, the
two-dimensional ground state manifold is well defined and separated
from the other states. By integrating over twist angles, one state
mixes with the other state when levels touch each other, therefore
in the Chern number evaluation, both levels participate and the
degenerate form of the Chern number should be used (Eq.
\ref{eq:degenerate}).

It is important to note that the degeneracy in the non-interacting
regime (Landau degeneracy) is fundamentally different from that of
the interacting hard-core case. In the non-interacting limit ($U \ll
J$), the degeneracy can be lifted by a local perturbation e.g. an
impurity, while in the hardcore case, the degeneracy remains in the
thermodynamic limit \cite{wen90}. The latter degeneracy is a
consequence of the global non-trivial properties of the manifold on
which the particles move rather than symmetries of the Hamiltonian
(e.g. the Ising model)\cite{wen95}. Recently, it was shown
\cite{senthil} that in presence of a gap, there is a direct
connection between the fractionalization and the topological
degeneracy. In particular, the amount of the degeneracy is related
to the statistics of the fractionalized quasiparticles e.g. in the
case of $\nu=1/2$, the two-fold degeneracy is related to 1/2 anyonic
statistics of the corresponding quasiparticles.

The ground state degeneracy prevents the direct integration of Eq.\
(\ref{eq:chern}) since wavefunctions would mix together when twist
angles vary. Therefore, one has to use Eq.\ (\ref{eq:degenerate})
and also resolve the extra gauge related to the ground state. We can
consider two possibilities: fixing the relative phase between the
two states in the ground state, or lifting the degeneracy by adding
some impurities. In the latter case, we can show that the system has
a topological order in spite of poor overlap with the Laughlin state
\cite{hafezi}. On the other hand, a significant amount of impurity
in the system may distort the energy spectrum, so that the
underlying physical properties of the lattice and fluxes could be
confounded by the artifacts due to the impurities, especially for
large $\alpha$. Therefore, in this letter we focus on the degenerate case.

\begin{figure}
  \centering%
  \begin{subfloat}%
\includegraphics[width=.45 \textwidth]{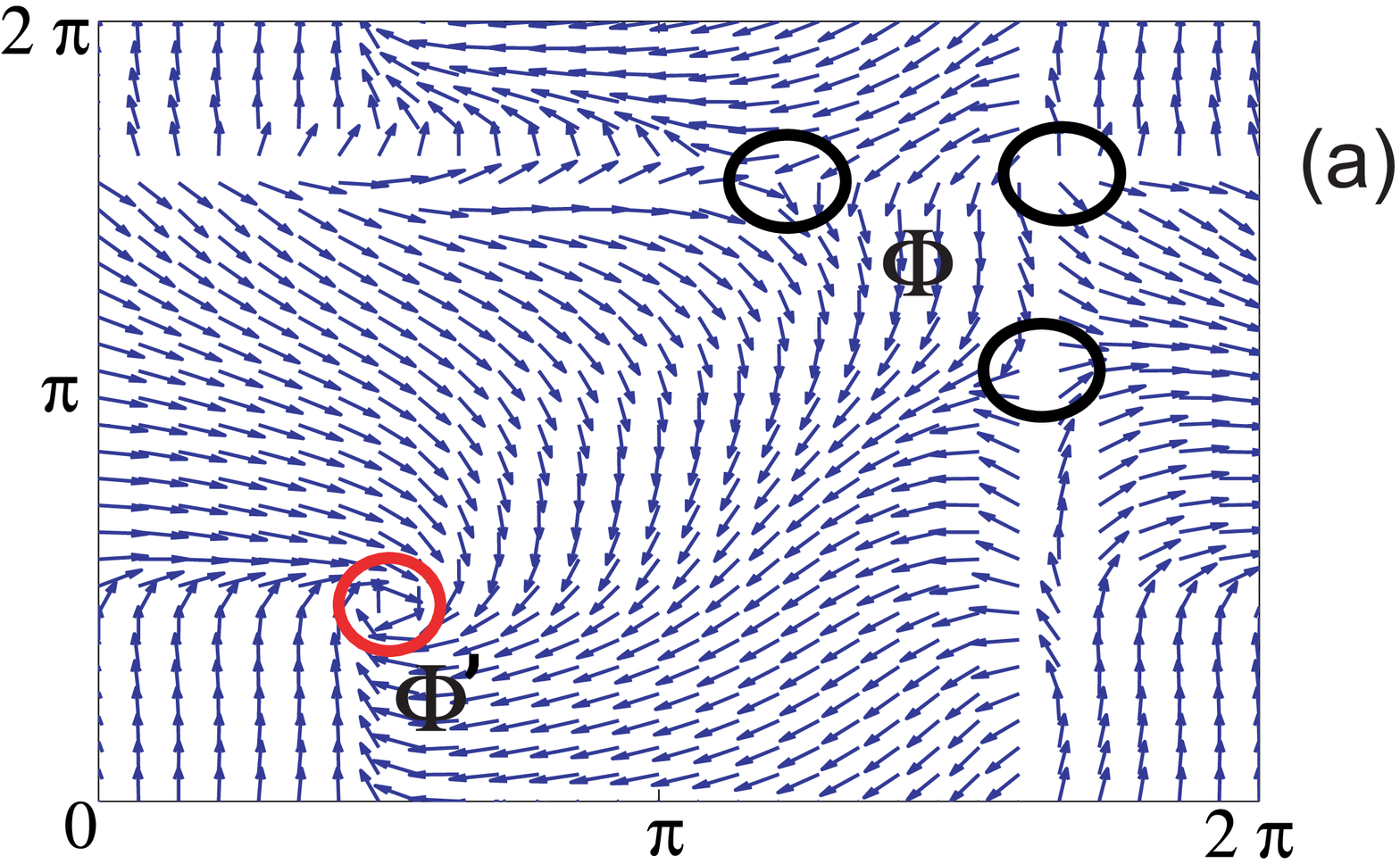}
 \end{subfloat}%
 \hspace{0cm}%
 \centering%
   \begin{subfloat}%
\includegraphics[width=.45 \textwidth]{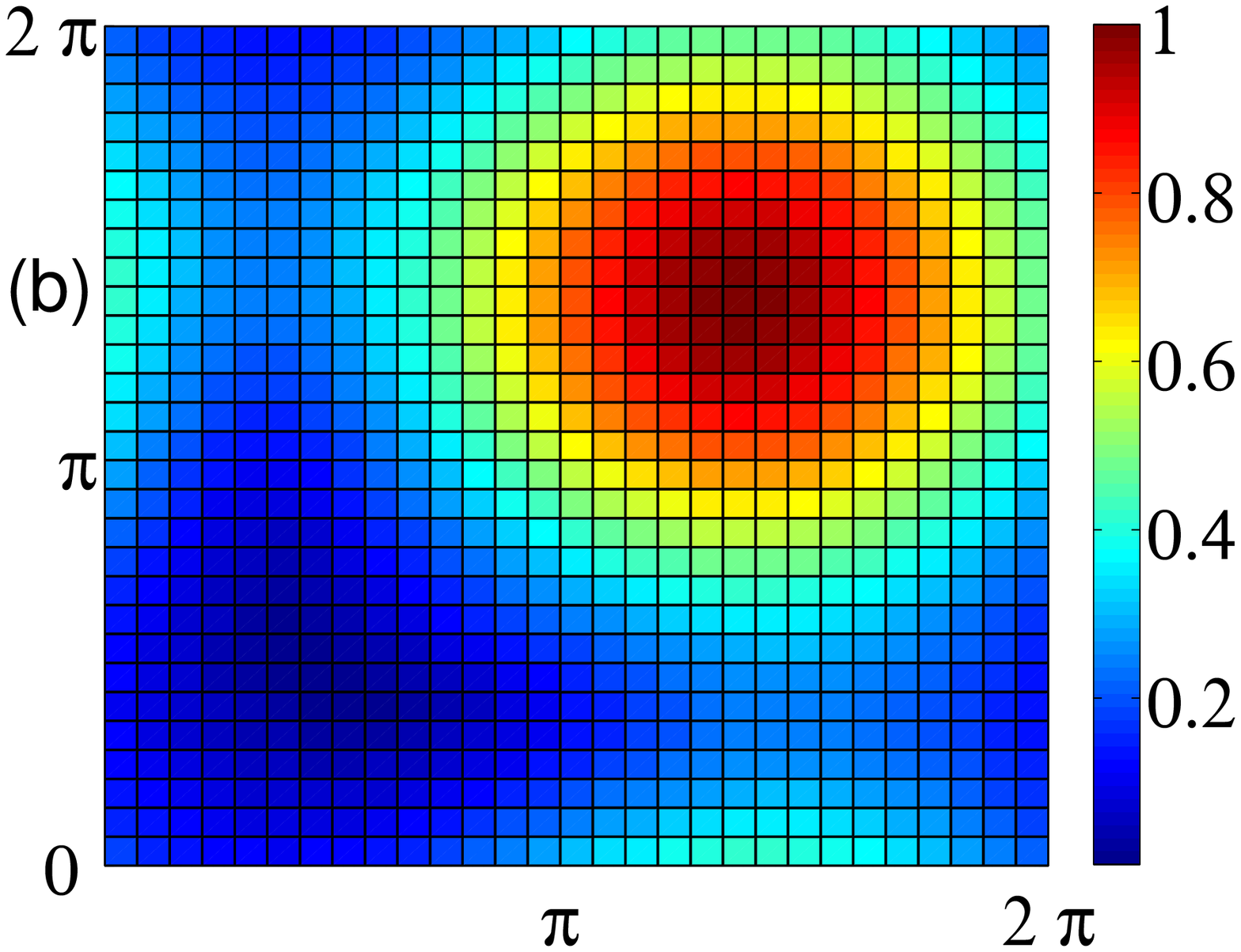}
 \end{subfloat}%
 \centering%
 \hspace{0cm}%
  \begin{subfloat}%
\includegraphics[width=.45 \textwidth]{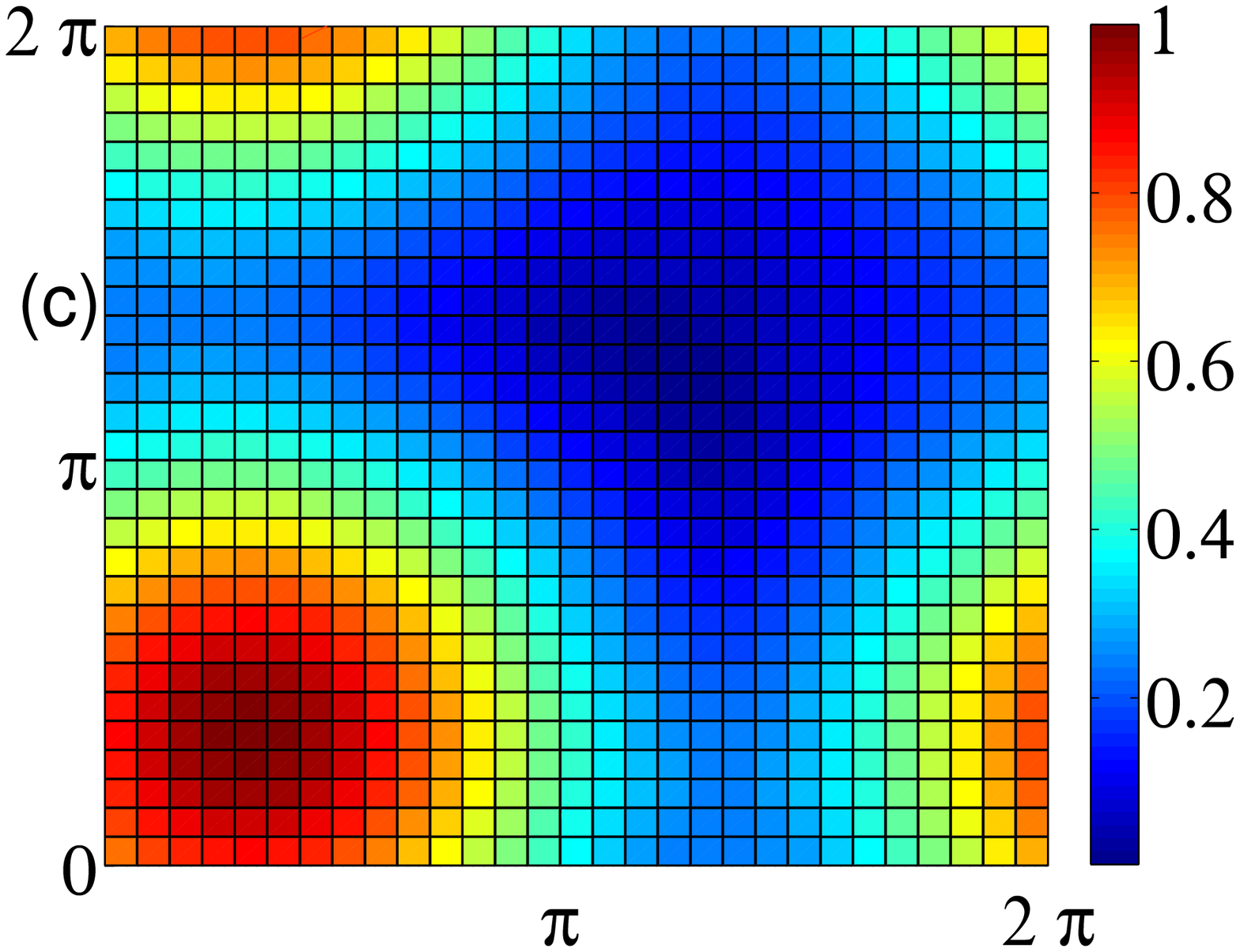}
  \end{subfloat}%

  \caption{(a) $\Omega(\theta_1, \theta_2)$ for fixed $\Phi$
and $\Phi'$. $\theta_1$ and $\theta_2$ changes form zero to $2\pi$.
This plot has been produced for 4 atoms in the hard-core limit on a
5x5 lattice ($\alpha=0.32$). (b) and (c): surface plots of $ {\rm
det} \Lambda_\Phi$ and $ {\rm det} \Lambda_\Phi'$ (blue is lower
than red). $\theta_1$ and $\theta_2$ changes form zero to $2\pi$.
The total vorticity corresponding to each of trial function ($\Phi$
or $\Phi'$) indicates a Chern number equal to one for the two
dimensional ground state manifold.}
  \label{fig:omega}
\end{figure}

We start with the simple case of a non-degenerate ground state on a
discrete ${\bf s}$-dimensional Hilbert space, $\Psi (\theta_1,
\theta_2)=(c_1, c_2,...,c_s)$. The one-dimensional gauge can be
resolved by making two conventions: in one convention the first
element and in the other the second element of the wavefunction in
the Hilbert space should be real i.e. we transform the ground state
$\Psi$ into $\Psi_{\Phi}= P \Phi=\Psi \Psi^\dagger \Phi$ where
$\Phi= (1,0,...,0)^{\dagger}$ is a ${\bf s}$-dimensional vector and
$P$ is a projection into the ground state and similarly with the
other reference vector $\Phi'= (0,1,...,0)^{\dagger}$. Hence, we can
uniquely determine the gauge $\chi$ which relates the two
corresponding vector fields: $e^{i \chi}=\Phi^\dagger P \Phi'$.
Therefore, the Chern number will be equal to the number of
vorticities of $\chi$ around regions where
$\Lambda_{\phi}=\Phi^\dagger P \Phi= |c_1|^2$ is zero.

For fixing the ${\bf q}$-dimensional ground state manifold gauge, we
take two reference  multiplets $\Phi$ and $\Phi'$ which are two
${\bf s} \times {\bf q}$ matrices (${\bf q=2}$ in our case). We
define an overlap matrix as $\Lambda_\phi=\Phi ^\dagger P \Phi$, and
consider the regions where ${\rm det}\ \Lambda_\Phi$ or ${\rm det}\
\Lambda'_\Phi$ vanishes (similar to zeros of the wave function in
the non-degenerate case). Hence, the Chern number for ${\bf q}$
degenerate states will be equal to the total winding number of
$\textrm{Tr}~ \chi^ {(\alpha,\beta)}$ for small neighborhoods $S_n$,
in which ${\rm det} \Lambda_\Phi$ vanishes. It should be noted that
the zeros of ${\rm det} \Lambda_\Phi$ and ${\rm det} \Lambda'_\Phi$
should not coincide in order to uniquely determine the total
vorticity. In our numerical calculation, we choose multiplets $\Phi$
and $\Phi'$ to be two sets of two degenerate ground states at two
different twist angles far apart e.g. $(0,0)$ and $(\pi,\pi)$. In
Fig.\ \ref{fig:omega}, we have plotted $\Omega={\rm
det}(\Phi^\dagger P \Phi')$, ${\rm det} \Lambda_\Phi$, and ${\rm
det} \Lambda_\Phi'$, found by numerical diagonalization of the
Hamiltonian over a grid ($30 \times 30$) of twist angles $\theta_1$
and $\theta_2$. The Chern number can be determined by counting the
number of vortices and it is readily seen that the winding number is
equal to one for the corresponding zeros of ${\rm det}
\Lambda_\Phi'$ and ${\rm det} \Lambda_\Phi$.

We have calculated the Chern number for fixed $\nu=1/2$ and
different $\alpha$'s by the method described above. The result is
shown in Tab.\ref{table:chern}. For $\alpha \ll 1$, we know from
previous calculation \cite{sorensen} that the ground state is the
Laughlin state and we expect to obtain a Chern number equal to 1/2
for each state i.e. total Chern number equal to one. For higher
$\alpha$, the lattice structure becomes more apparent and the
overlap with the Laughlin state decreases rapidly. However, in our
calculation, the ground state remains two-fold degenerate and the
associated Chern number remains equal to one before reaching some
critical $\alpha_c \simeq 0.4$. Hence, we expect to have similar
topological order and fractional statistics of the excitations on
the lattice in this regime.

For higher flux densities, $\alpha > \alpha_c$, the two-fold ground
state degeneracy is no longer valid everywhere on the torus of the
boundary condition. In this regime, the issue of degeneracy is more
subtle, and finite size  effect becomes significant. The
translational symmetry argument \cite{haldane85} is no longer valid
and the degeneracy of the ground state varies periodically with the
system size \cite{Kol}. Some gaps might be due to the finite size
and vanish in the thermodynamic limit. To investigate this, we study
the ground state degeneracy as a function of boundary angles
($\theta_1, \theta_2$) which are not physical observable. Therefore,
the degeneracy in thermodynamic limit should  not depend on the
their value. In particular, Fig. \ref{fig:energy-levels}b shows the
energy levels of five particles at $\alpha=0.4$ for different values
of twist angles. The first and the second level are split at
($\theta_1=\theta_2=0$), while they touch each other at
($\theta_1=\theta_2=\pi$). We have observed similar behavior  for
different number of particles and lattice sizes e.g. 3 and 4 atoms
at $\alpha=0.5$. Therefore, the ground state enters a different
regime which is a subject for further investigation. Existence of
the topological order does not require a very strong interaction
i.e. hard-core limit. Even at finite interaction strength $U \sim J
\alpha $, we have observed the same topological order with the help
of the Chern number calculation. If $U$ gets further smaller, the
energy gap above the ground state diminishes \cite{hafezi} and the
topological order disappears.

\begin{table}[t]
\begin{center}
\begin{tabular}{|c|c|c|c|c|c|c|c|}   \hline
{\em Atoms  }      & {\em Lattice } &      {\em $\alpha$}  & gap /J & {\em Chern/state}& Overlap  \\
\hline  3           &    6x6  &    .17      &    0.24  &    1/2 &  0.99 \\
\hline  4           &    6x6  &    .22      &    0.24   &    1/2 &  0.98 \\
\hline  3           &    5x5  &    .24      &    0.23     &1/2 & 0.98 \\
\hline  3           &    4x5  &    .3        &   0.18     &1/2 & 0.91 \\
\hline  4           &    5x5 &    .32       &    0.15  &    1/2 & 0.78\\
\hline  3           &    4x4 &    .375     &  0.03   & 1/2 & 0.29\\
\hline
\end{tabular}
\end{center}
\caption{Chern Number for different configurations in the hard-core
limit for fixed filling factor $\nu=1/2$. The Laughlin state overlap
is shown in the last column. although it deviates from the Laughlin
state. Although the ground state deviates from the Laughlin state,
the Chern number remains equal to one half per state before reaching
some critical $\alpha_c \simeq 0.4$ where the energy gap vanishes. }
\label{table:chern}
\end{table}

One of the impediment of the experimental realization of Quantum
Hall state is the smallness of the gap which can be improved in
presence of the dipole-dipole interaction \cite{hafezi}. The dipole
interaction can be represented as extra term $\sum_{ij}U_d\ {\bf
n_i}\ {\bf n_j} /|{\bf r_i}-{\bf r_j}|^3$ in the Hamiltonian Eq.\
(\ref{eq:Hamiltonian}), where $n_i$ is the number of particles at
location ${\bf r_i}$ in the units of lattice spacing and $U_d$ is
the strength of the interaction. The magnetic dipole-dipole
interaction has been achieved in Bose-Einstein condensation of
Chromium \cite{griesmaier}, however, for a lattice realization,
polar molecules with strong permanent electric dipole moments are
more promising candidates, where the dipole interaction can be an
order of magnitude greater than the tunneling energy. In the
presence of such strong long-range interaction, the ground state deviates from
the conventional FQH state even in the continuum case (i.e. for even
$\alpha < 0.2 $, the overlap with the Laughlin wavefunction
decreases by increasing the strength of the dipole interaction).
However, by evaluating Chern number, we are able to identify the
topological order of the system, that turns out to be intact i.e.
Chern number equal to one for the two-fold degenerate ground state.

In conclusion, we have investigated a method to unambiguously
calculate the Chern number for the ground state of a system.  For
the FQHE system on a lattice that we have investigated, the Laughlin
wavefunction ceases to be a good description of the ground state for
high fluxes $\alpha\gtrsim 0.25$, but the Chern number remains 1/2
per state until $\alpha \lesssim 0.4$ which is a direct indication
of topological order in the system. Calculating Chern numbers by
this method can be generalized for finite lattice systems to
properly characterize the ground state manifold which is otherwise
impossible with conventional overlap methods.

We thank K. Yang and S. Girvin for fruitful discussions. This work
was partially supported by the NSF Career award, Packard Foundation,
AFSOR and the Danish Natural Science Research Council.


\end{document}